# Scale-robust Low Resistance Transport in Atomic Layer Deposited Topological Semimetal Wafers on Amorphous Substrate

Dong-Hyun Lim[1,#], Young-Min Song[1,#], Yeji Kim[1], Ae Rim Choi[1], Hyun-Mi Kim[2], Hyeongkeun Kim[2], Sujin Kwon[3], Bonggeun Shong[3], Justin Shih[4], Asir Intisar Khan[4,*], and Il-Kwon Oh[1,5*]

[1]*Department of Intelligence Semiconductor Engineering, Ajou University, Suwon 16499, Republic of Korea*
[2]*Korea Electronics Technology Institute, Seongnam-si, Republic of Korea*
[3]*Department of Chemical Engineering, Hongik University, Republic of Korea*
[4]*Department of Electrical Engineering and Computer Sciences, University of California, Berkeley, CA 94720, USA*
[5]*Department of Electrical and Computer Engineering, Ajou University, Suwon 16499, Republic of Korea*

*Corresponding Author E-mails: asir@berkeley.edu, ikoh@ajou.ac.kr
[#] These authors are equally contributed.

## Abstract

As data-centric computing advances, energy-efficient interconnects are increasingly critical for AI-driven systems. Traditional metal conductors face severe limitations at nanoscale due to increased resistivity from surface scattering. In response, this study demonstrates the first wafer-scale realization of an amorphous topological semimetal, tantalum phosphide (TaP), grown directly on amorphous $SiO_2$ substrates (without any seed layers) using low-temperature atomic layer deposition (ALD). The resulting TaP films exhibit unconventional resistivity scaling: decreasing resistivity with decreasing thickness: reaching 227 μΩ·cm at ~2.3 nm film thickness. This behavior, observed without crystalline order or seed layers, indicates dominant surface conduction and establishes ALD-TaP as a promising candidate for back-end-of-line integration. The films also show excellent conformality, stoichiometry control, and thermal stability up to 600 °C. A two-channel conduction model confirms surface-dominated transport in ultrathin regimes, further supported by enhanced conductivity in multi-stacked configurations. These findings highlight the potential of amorphous topological semimetals for future high-density, low-power electronic interconnects and expand the applicability of ALD for integrating novel quantum materials at scale.

**Introduction**

With the rise of data-centric computing, energy efficiency is a critical challenge for AI-driven applications. The push for both energy savings and higher performance has driven semiconductor scaling into new territory. As technologies evolve, approaches such as 3D integration of logic and memory promise greater computing density and speed. Yet even as computing and memory units advance through novel materials and architectures, the nanoscale metal interconnects linking them pose fundamental limitations. At reduced dimensions, these metal conductors suffer increased electron–surface scattering, leading to higher resistivity, and ultimately, greater power dissipation and signal delay. The need for nanoscale, low-resistance conductors is therefore critical for the hyper-scale electronics era.

To address this need, topological Weyl semimetals (TSMs), particularly transition metal pnictides such as NbAs, NbP, TaP and TaAs (*1–5*) have emerged as promising candidates due to their ability to conduct through surface states that are more resilient to scattering than those in conventional metals. As the film thickness decreases, these surface channels can dominate transport, leading to improved conductivity. For example, recent studies on crystalline NbAs nanobelts and nanowires (*6, 7*) have shown nearly an order of magnitude lower resistivity than their bulk counterparts, indicating a surface-driven conduction mechanism. However, realizing such behavior in crystalline NbAs required high-temperature (>700 °C) growth, which is incompatible with the thermal limits of modern back-end-of-line (BEOL) semiconductor processing. The topological nature of TaP has been confirmed via angle-resolved photoemission spectroscopy (ARPES) on single crystals, revealing clear Weyl nodes and Fermi arcs. (*8*) However, most synthetic efforts have focused on bulk or epitaxial growth at (>500 °C), limiting BEOL compatibility. (*8–10*)

On the other hand, recent work showed reduced electrical resistivity and surface-dominated conduction in NbP/Nb bilayer thin films, where nanocrystalline NbP was sputter-deposited (at 400 °C) on an epitaxial 4 nm Nb seed. (*11*) However, it remains unclear whether the observed low resistance is an intrinsically surface-dominated property of topological semimetal NbP, or influenced by the Nb/NbP interface, particularly given the presence of a crystalline seed layer. The additional seed layer thickness also imposes scaling limitations, and physical vapor deposition lacks the conformality required for complex device geometries in hyper-scaled electronics. A key question thus arises: can similar transport behavior be achieved directly on amorphous substrates, without a crystalline seed and under BEOL-compatible conditions, especially in high-density device architectures that demand extreme scaling and conformal integration?

To address these, here we demonstrate the first wafer-scale realization of a topological semimetal, TaP, on an amorphous SiO$_2$ substrate using atomic layer deposition (ALD) at 170 °C -without the need for a crystalline epitaxial seed. The resulting amorphous TaP films exhibit an unconventional thickness–resistivity relationship, with resistivity decreasing as the film is thinned to the nanometer regime (e.g., from ~1000 μΩ·cm at 18 nm to ~227 μΩ·cm at 2.3 nm). This confirms scale-robust, low-resistance transport even in the absence of long- or short-range crystalline order or epitaxial interfaces. In addition, the ALD process offers precise thickness control and excellent conformality, enabling uniform coating in complex, high-aspect-ratio structures and underscoring its compatibility with BEOL integration.

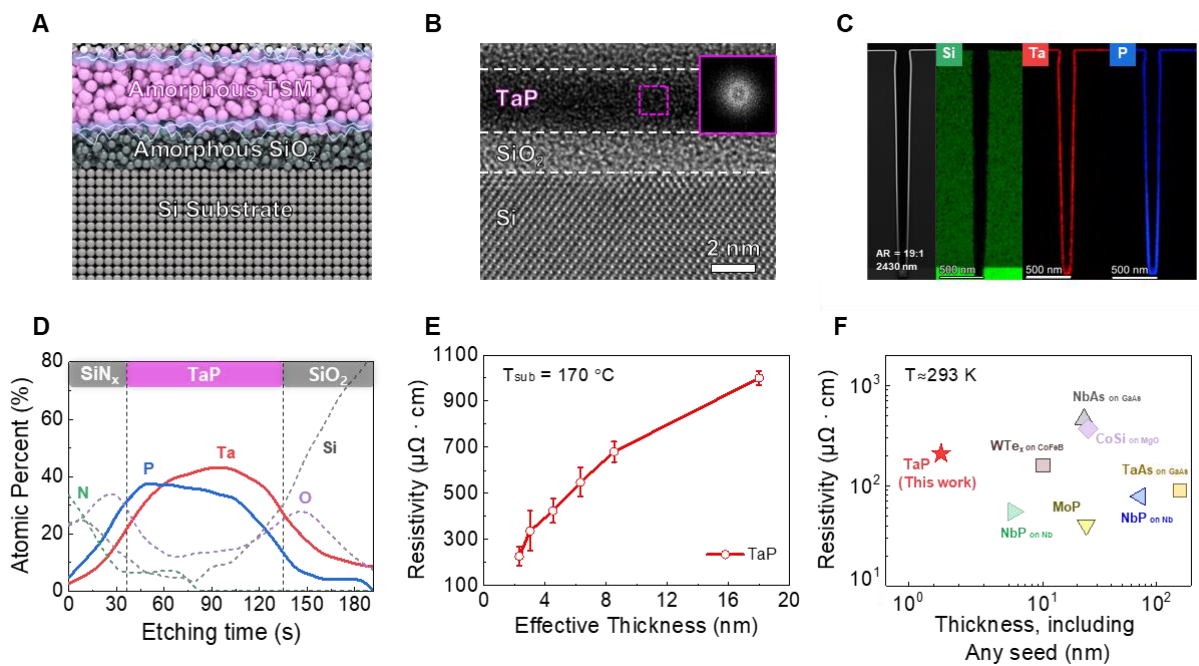

**Figure 1. Wafer-scale ALD of TaP on an amorphous substrate and unconventional resistivity scaling.** (**A**) Schematic of an amorphous TaP film deposited directly on amorphous SiO$_2$ atop a silicon substrate with SiO$_2$. (**B**) Cross-sectional HR-TEM image of a ultrathin amorphous TaP film deposited by ALD on SiO$_2$/Si substrate. The film is capped in situ with plasma-enhanced ALD SiN$_x$. Inset image shows a SAED pattern with diffuse rings. (**C**) STEM HAADF image of representative trench structures with aspect ratio (AR) ~19:1 and depth ~2.4 μm. Cross-sectional EDS elemental maps of Si, Ta, and P across a high-aspect-ratio trench structure coated with ALD TaP. (**D**) Depth-resolved XPS of a SiN$_x$/TaP/SiO$_x$/Si stack, displaying Ta and P with near 1:1 atomic ratio across the TaP layer. (**E**) Room-temperature resistivity of ALD TaP films deposited on SiO$_2$/Si at 170°C as a function of a 'effective' thickness. The 'effective' thickness is defined as the electrically active TaP region after excluding the ~1.5 nm SiN$_x$/TaP interfacial layer formed during the N$_2$ plasma SiN$_x$ capping process, consistent with prior approaches for dead-layer correction in ultrathin conductors (*12*). Based on this definition, the thinnest TaP film investigated in this study has an effective thickness of 2.3 nm, for which the resistivity is measured as 227 ± 41 μΩ·cm. (**F**) Room temperature resistivity versus thickness including any seed layer for various materials. Here, our ALD-grown TaP semimetal resistivity is shown together with TSMs (CoSi, NbP, MoP, WTe$_x$, and NbAS) and topological insulator (Bi$_2$Se$_3$). (*2*, *11*, *13–18*) ALD-grown TaP shows one of the lowest resistivity in ultrathin films.

TaP thin films were deposited directly on amorphous SiO$_2$/Si substrates using plasma-enhanced ALD at 170 °C, and capped in-situ with ALD Si$_3$N$_x$ to prevent oxidation (schematic in **Fig. 1A**). The resulting TaP films are as-deposited amorphous which we imaged using high-resolution transmission electron microscopy (HR-TEM). Cross-sectional HR-TEM of an ultrathin TaP film (**Fig. 1B**) reveals a uniform amorphous contrast with no short or long-range crystalline ordering, which is further confirmed by the corresponding selected-area electron diffraction (SAED) pattern displaying diffuse rings. The process retains the inherent conformality of ALD at the wafer scale. Uniform TaP deposition was achieved on 150 mm wafers patterned with trench arrays ~2.4 μm deep and with aspect ratios up to 19:1, as confirmed using scanning transmission electron microscopy (STEM) in high-angle annular dark-field (HAADF) mode (**Fig. 1C**). Depth-resolved X-ray photoelectron spectroscopy (XPS) analysis (**Fig. 1D**) indicates uniform Ta and P incorporation with a near 1:1 atomic ratio throughout the film thickness. These results show that TaP can be stabilized in ultrathin form on amorphous substrates, without reliance on crystalline templates or metallic seed layers: an important requirement for BEOL integration, where conductive interconnect layers need to be deposited conformally on amorphous dielectric surfaces.

Electrical transport measurements (**Fig. 1E**) show that the room temperature resistivity of amorphous TaP thin films decreases by nearly an order of magnitude as the thickness is scaled from 18 nm down to ~2.3 nm. This unconventional resistivity scaling trend can be understood as a progressive dominance of surface channels at reduced dimensions, which could account for the lower resistivity observed in thinner films. Similar behavior has been reported in transition-metal pnictide topological semimetals such as NbAs and NbP (*16*, *17*), where surface-dominated conduction has been attributed to play an important role in both crystalline and nanocrystalline forms. More intriguingly, recent study of amorphous Bi$_2$Se$_3$ demonstrates that topologically protected surface states can persist even without long-range crystalline order: an indication that surface transport may be robust to structural disorder. (*19*) Complementary evidence of disorder-tolerant surface transport has also been reported in amorphous CoSi thin films, where an unconventional decrease in resistivity with decreasing thickness was observed. (*13*) To this end, the unconventional scaling trend observed in amorphous TaP suggests that surface-dominated conduction mechanisms may likewise contribute here, a possibility we revisit later below.

**Fig. 1F** further shows that at the ultrathin effective thickness limit of ~2.3 nm, the room-temperature resistivity of amorphous TaP films reaches 227 ± 41 μΩ·cm; among the lowest reports for sub-5 nm thin conductive thin films. Among topological semimetals, our recent report on NbP grown on a crystalline 4 nm Nb seed (NbP/Nb bilayer stack) achieved ~55 μΩ·cm resistivity at a total thickness of 5.5 nm. In contrast, seed-free ALD TaP in this work achieves a resistivity of 227 ± 41 μΩ·cm at an effective thickness of only ~2.3 nm, underscoring its promise as an ultrathin conductive layer despite its fully amorphous structure.

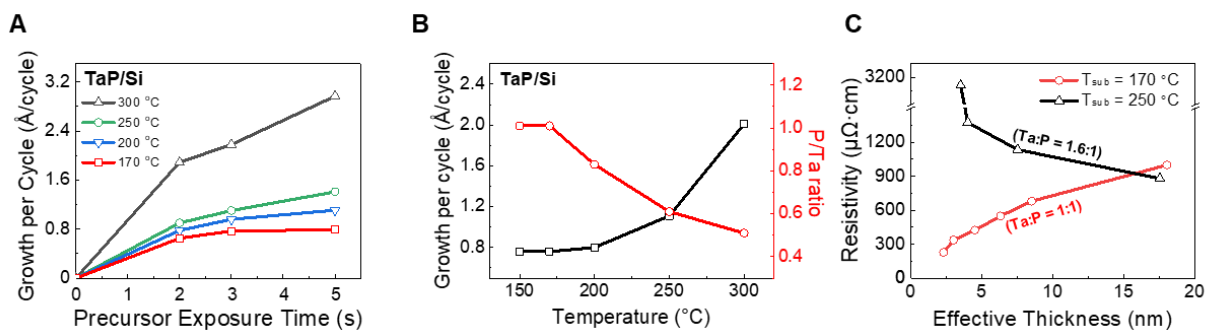

**Figure 2. Advanced phosphide ALD process for stoichiometric TaP growth.** (**A**) Growth per cycle (GPC) behavior versus precursor exposure at various deposition temperatures. (**B**) GPC and P/Ta ratio as a function of deposition temperature, showing transition to Ta-rich films above 250 °C. (**C**) Thickness-dependent resistivity of films grown at 170 °C (stoichiometric) versus 250 °C (Ta-rich), highlighting that only stoichiometric films exhibit an unconventional resistivity scaling, where resistivity decreases with decreasing thickness.

TaP ALD process is optimized at 170 °C, showing self-limiting growth (**Fig. 2A**), with GPC saturating at ~0.8 Å/cycle and film thickness. At higher temperatures (250–300 °C), the GPC nearly doubles and loses saturation. Correspondingly, the Ta:P ratio shifts from ~1:1 at 170 °C to ~2:1 at 300 °C (**Fig. 2B**), yielding Ta-rich films. The stoichiometric TaP films (deposited at 170 °C) exhibit unconventional resistivity scaling (**Fig. 2C**) trend. In contrast, Ta-rich ALD films (250 °C) display conventional metallic scaling with increasing resistivity at reduced thickness. These results show that maintaining stoichiometry is essential to realize the low resistivity in thinner films expected for topological semimetals. Importantly, the effect is observed in our seed-free amorphous TaP ALD films, demonstrating a BEOL-compatible route for integrating such materials into future nanoscale interconnect architectures.

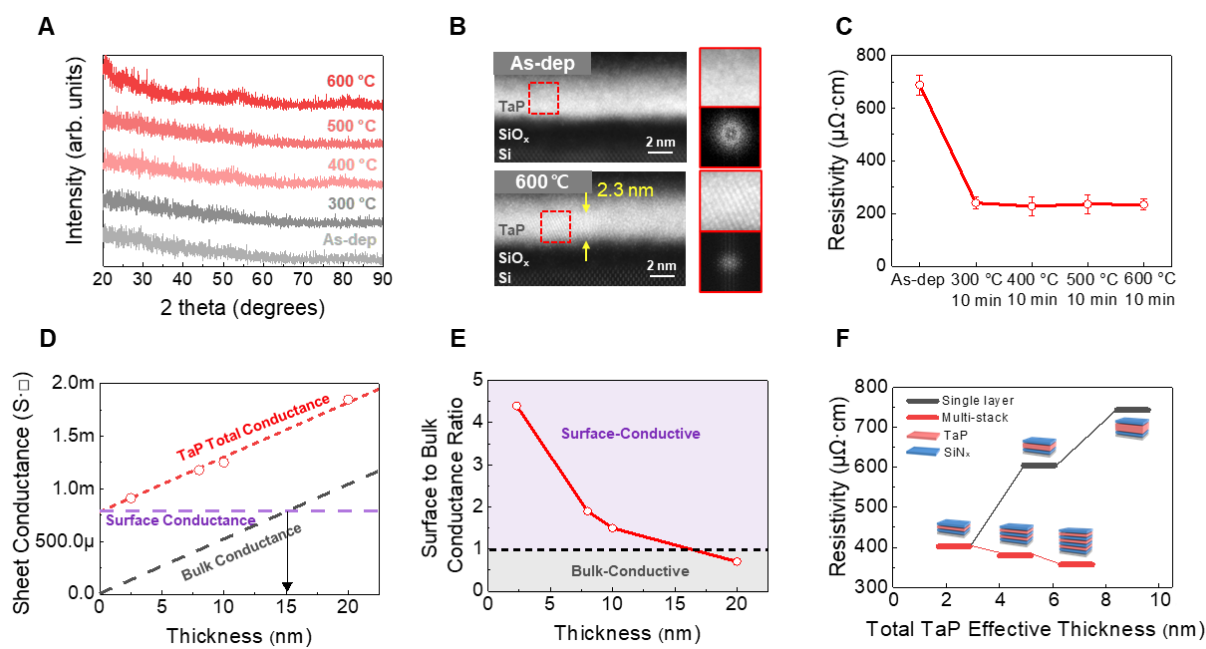

**Figure 3. Microstructure, electrical stability and surface conduction of ALD TaP films. (A–C)** Structural and electrical stability of ultrathin TaP films upon annealing. **(A)** XRD patterns of TaP films annealed at 300–600 °C, showing an amorphous structure up to 500 °C and a weak (110) tetragonal peak appearing at 600 °C. **(B)** Cross-sectional HAADF–STEM images of as-deposited and 600 °C-annealed TaP films on SiO$_x$/Si, together with the corresponding FFT patterns showing an amorphous-to-nanocrystalline transition. When viewed along the TaP [02$\bar{1}$] zone axis, inverse FFTs taken from the nanocrystalline regions reveal lattice fringes along the [100] direction with a real-space periodicity of ~3.35 Å, corresponding to the a-axis lattice parameter of tetragonal TaP (bulk a = 3.32 Å). **(C)** Temperature-dependent resistivity of ultrathin TaP films, which decreases after 300 °C annealing and remains nearly constant (~220 μΩ·cm) up to 600 °C annealing temperature, demonstrating thermal stability. **(D–F)** Thickness- and configuration-dependent conduction behavior of TaP films. **(D)** Total conductance versus film thickness fitted using a two-channel (bulk + surface) conduction model assuming thickness-independent surface conductance. (*11*) **(E)** Ratio of surface-to-bulk conductance showing that films thinner than ~15 nm are dominated by surface transport. **(F)** Comparison of resistivity between multi-stacked and single-layer TaP films of equivalent total thickness. The multi-stacked configuration exhibits lower resistivity and a decreasing trend with increasing total thickness, in contrast to the single-layer case.

The structural and electrical properties of ALD TaP films demonstrate thermal robustness up to 600 °C (**Fig. 3A–C**). X-ray diffraction (XRD) patterns of 2.3 nm TaP films show no discernible diffraction peaks up to 500 °C, indicating an amorphous structure, while a weak (110) reflection appears at 600 °C, signifying limited nano-crystallization without phase transformation. Cross-sectional HAADF–STEM and FFT analyses (**Fig. 3B**) confirm this amorphous-to-nanocrystalline transition and reveal no interfacial reaction or interdiffusion at the TaP/SiO$_2$ boundary. Consistently, the resistivity (**Fig. 3C**) decreases sharply from ~700 μΩ·cm in the as-deposited state to ~220 μΩ·cm after annealing at 300 °C and remains nearly unchanged upon further annealing up to 600 °C. These results confirm the simultaneous structural and electrical stability of ALD TaP under back-end-of-the-line compatible thermal budgets.

To elucidate the conduction mechanism, the thickness-dependent conductance was analyzed using a two-channel (surface + bulk) model (**Fig. 3D,E**). The total conductance increases linearly with thickness, whereas the extracted surface component remains constant, indicating a parallel contribution from bulk and surface channels. The surface-to-bulk conductance ratio exceeds unity for films thinner than ~15 nm, identifying a surface-dominated transport regime. The persistence of high conductivity in amorphous TaP suggests that conduction is mediated by robust surface states or localized bonding configurations, largely independent of long-range crystallinity. To further probe the role of surface conduction, we compared the resistivity of single-layer and multi-stacked TaP configurations with equivalent total thicknesses (**Fig. 3F**). While the resistivity of single-layer films increases with thickness, the multi-stacked architecture exhibits the opposite trend: a monotonic decrease in resistivity with increasing total TaP thickness. This contrasting behavior indicates that each TaP/SiN$_x$ interface contributes a high-conductivity surface channel, and that serial stacking effectively enhances overall conductivity by introducing additional conducting interfaces.

**Discussion and Outlook**

In summary, our study demonstrates the first demonstration of controllable ALD route to stoichiometric TaP; an achievement that expands ALD chemistry beyond conventional nitrides and oxides to include phosphorus-based refractory compounds. The process enables conformal, thickness-controlled growth of ultrathin semi-metallic films while preserving thermal and electrical stability, therefore enabling a scalable synthesis route of emerging topological semimetals. Compared with other synthesis routes, ALD offers precise thickness control, substrate conformity, and compatibility with back-end thermal budgets, positioning it as a practical pathway for integrating quantum and correlated materials into nanoscale architectures. Beyond the process advance, our results reveal robust surface conduction in amorphous TaP, underscoring that non-crystalline topological semimetal analogues can retain key electronic functionalities including low resistivity at ultra-thin films.

**Materials and Methods**

TaP thin films were deposited by plasma-enhanced ALD (PE-ALD) in a 6-inch showerhead-type reactor equipped with a capacitively coupled plasma (CCP) source. Substrate temperatures were varied from 150–300 °C, with 170 °C identified as the optimal condition for stoichiometric growth. Immediately after deposition, all films were *in-situ* capped with $SiN_x$ layer grown by ALD to prevent surface oxidation.

Film thickness was measured by spectroscopic ellipsometry (Elli-SE, Ellipso Technology Co., Ltd.). Composition was determined by XPS (Nexsa, Thermo Fisher Scientific) including depth profiling. Depth profiles were obtained using a 1 keV $Ar^+$ ion beam with 8 s etch intervals. Crystal structure was examined by grazing-incidence XRD (SmartLab 9 kW, Rigaku, Cu Kα radiation, 0.5° incidence), and conformality was tested on Si trench structures with 2.43 µm depth and aspect ratio of 19:1. High-resolution imaging was performed using a double Cs-corrected TEM (Themis Z, ThermoFisher Scientific) with 80 pm resolution, operated at 300 kV. Structural analysis was conducted by atomic-resolution HAADF-STEM with a convergence semi-angle of 17.9 mrad and detector collection angles of 62–200 mrad, providing Z-contrast imaging with minimal delocalization. Compositional analysis of the thin film was carried out by energy dispersive spectroscopy (EDS) in the same instrument using four windowless detectors (SuperXG2). The combined HAADF-STEM and EDS measurements enabled direct correlation of structural and compositional information at the atomic scale.

Electrical properties were evaluated on $SiN_x$/TaP/Si stacks using a contactless eddy-current system (EddyCus® TF lab 2020, SURAGUS). Sheet resistance of the $SiN_x$/TaP/Si stack ($R_{total}$) was measured by the non-contact eddy-current technique. The intrinsic TaP sheet resistance ($R_{TaP}$) was obtained by subtracting the parallel contribution from the p-type Si substrate according to:

$$\frac{1}{R_{TaP}} = \frac{1}{R_{Total}} - \frac{1}{R_{Si}}$$

where $R_{Si}$ is the independently measured sheet resistance of the bare Si wafer (~150 Ω/□). The TaP resistivity ($\rho_{TaP}$) was obtained as the product of the TaP film thickness ($t_{TaP}$) and the extracted sheet resistance ($R_{TaP}$). For multilayer TaP/$SiN_x$ stacks, the total sheet resistance was extracted using a parallel-resistance model, assuming each TaP layer contributes independently. The corresponding resistivity was calculated from the total stack thickness and extracted sheet resistance.

**Acknowledgement**

This work was supported by the National Research Foundation of Korea (NRF) grant funded by the Korea government (MSIT)(RS-2024-00357895). H.M. Kim and H. Kim was supported by Material Innovation Leading Project through the National Research Foundation of Korea (NRF) funded by the Ministry of Science and ICT (2020M3H4A3081879). This work was supported by the Ajou University research fund. This work was supported by K-CHIPS(Korea Collaborative & High-tech Initiative for Prospective Semiconductor Research) (2410012153, RS-2025-02310666, 25073-15FC) funded by the Ministry of Trade, Industry & Energy (MOTIE, Korea). This research was supported by the Commercialization Promotion Agency for R&D Outcomes(COMPA) funded by the Ministry of Science and ICT(MSIT)(2710084665).